\documentclass[english,aps,pra,showpacs,twocolumn,floatfix]{revtex4}

\usepackage{amsmath}
\usepackage{amssymb}
\usepackage{amsfonts}
\usepackage{mathrsfs}
\usepackage{epsfig}
\usepackage{grffile}
\usepackage{bbm}
\usepackage{babel}

\newcommand{\ud}{\mathrm{d}}
\renewcommand{\Re}{\operatorname{Re}}

\newcommand{\Tr}{\operatorname{Tr}}
\newcommand{\tot}{\text{tot}}

\newcommand{\eff}{\text{eff}}
\newcommand{\ket}[1]{| #1 \rangle}
\newcommand{\bra}[1]{\langle #1 |}
\newcommand{\vsigma}{{\boldsymbol{\sigma}}}
\newcommand{\mc}[1]{\mathcal{#1}}
\newcommand{\hmc}[1]{\hat{\mathcal{#1}}}
\newcommand{\tmc}[1]{\tilde{\mathcal{#1}}}

\newcommand{\Id}{\mathbbm{1}}

\newcommand{\tmat}[1] {\begin{bmatrix}#1\end{bmatrix}}
\newcommand{\tsmat}[1] {\left[\begin{smallmatrix}#1\end{smallmatrix}\right]}

\usepackage{color}
\usepackage[draft]{hyperref}

\newcommand{\lmu} {Department of Physics and Arnold Sommerfeld Center for Theoretical Physics,
Ludwig-Maximilians-Universit{\"a}t M{\"u}nchen, Theresienstr.\ 37, 80333 Munich, Germany}

\newcommand{\Title} {Algebraic versus exponential decoherence in dissipative many-particle systems}
\newcommand{\Authors}
{
\author{Zi Cai}
\affiliation{\lmu}
\author{Thomas Barthel}
\affiliation{\lmu}
}
\newcommand{\Date} {March 30, 2013}

\begin{document}

\title{\Title}
\Authors

\date{\Date}

\begin{abstract}
The interplay between dissipation and internal interactions in quantum many-body systems gives rise to a wealth of novel phenomena. Here we investigate spin-$1/2$ chains with uniform local couplings to a Markovian environment using the time-dependent density matrix renormalization group (tDMRG). For the open XXZ model, we discover that the decoherence time diverges in the thermodynamic limit. The coherence decay is then algebraic instead of exponential. This is due to a vanishing gap in the spectrum of the corresponding Liouville superoperator and can be explained on the basis of a perturbative treatment. In contrast, decoherence in the open transverse-field Ising model is found to be always exponential. In this case, the internal interactions can both facilitate and impede the environment-induced decoherence.
\end{abstract}


\maketitle

\section{Introduction}
Every quantum system that we wish to study or model is inevitably coupled to some form of environment and hence an \emph{open} quantum system \cite{Weiss1999,Gardiner2004,Breuer2007,Alicki2007,Rivas2012}. The coupling to the environment can for example induce decay of quantum coherence (decoherence) and dissipation. To take account of these effects is particularly interesting and complex when the system itself is already an interacting many-body system. Recently, first theoretical \cite{Cazalilla2006,Prosen2008,Daley2009,Diehl2010,Torre2010,Prosen2011b,Karevski2013-110,Schwager2012,Poletti2012,Poletti2012a,Bernier2012,Honing2012,Horstmann2013,Arenas2013,Lorenzo2013-87} and experimental \cite{Barreiro2010,Barreiro2011,Schindler2012} investigations were presented for this scenario. Whereas decoherence is usually seen as an obstacle for quantum simulation \cite{Buluta2009-326} and information processing \cite{Nielsen2000,Jones2012}, it has also been suggested that one could exploit the effect for the preparation of desired many-body states by engineering the dissipative processes \cite{Diehl2008,Verstraete2009,Diehl2010a,Diehl2011b}.

For \emph{Markovian} environments \cite{Lindblad1976-48,Gorini1976-17,Wolf2008-279}, the system state $\rho$ evolves according to the Lindblad master equation $\partial_t \rho(t) = \hmc{L}\rho(t)$. The decoherence behavior is determined by the spectral gap of the Liouville superoperator $\hmc{L}$. For the textbook-type scenarios of finite-size systems or many-body systems without interaction, the gap is necessarily finite and, consequently, quantum coherence decays exponentially with time \cite{Weiss1999,Gardiner2004,Breuer2007,Alicki2007}. This imposes strong limitations for many quantum simulation and information processing applications. In this work, we find however that Markovianity does \emph{not} necessarily imply exponential decoherence. For cases where the system itself is a many-body system with internal interactions, we show that the Liouvillian gap can close in the thermodynamic limit and lead to a divergent decoherence time due to an interplay of dissipation and interaction. The coherence decay then becomes algebraic, i.e., follows a power law, instead of being exponential. This novel phenomenon is reminiscent of the importance of the Hamiltonian gap for closed many-body systems which is intimately related to quantum phase transitions \cite{Sachdev1999}, the scaling behavior of entanglement, and the spatial decay of quantum correlations \cite{Latorre2009,Eisert2008}.
\begin{figure}[b]
\centering
\includegraphics[width=1\linewidth]{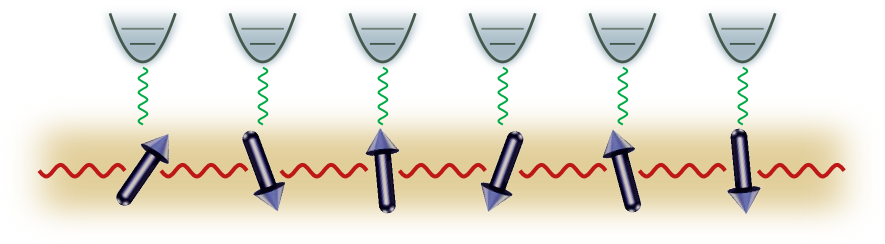}
\caption{A quantum spin chain uniformly coupled to the environment via the $z$-components of the spins.}
\end{figure}

Specifically, let us consider spin-$1/2$ lattice systems with local couplings to a Markovian environment. The Lindblad master equation \cite{Lindblad1976-48,Gorini1976-17,Wolf2008-279} reads ($\hbar$=1)
\begin{eqnarray} \nonumber
	\partial_t{\rho} &=& \hmc{L}\rho=\hmc{H}\rho+\hmc{D}\rho\\
	                 &=&-i[H,\rho]+\gamma\sum_i \Big(L_i \rho L_i^\dag-\frac{1}{2} \{L_i^\dag L_i,\rho\}\Big).  \label{eq:master}
\end{eqnarray}
The Liouville superoperator $\hmc{L}$ contains two parts: $\hmc{H}\rho=-i[H,\rho]$ generates the evolution due to the system Hamiltonian $H$ while the dissipative process is described by $\hmc{D}\rho$. This equation of motion describes, for example, systems in the weak-coupling regime (Born-Markov-secular approximation), singular-coupling regime, or the time average of a system with stochastic Hamiltonian terms as described in appendix~\ref{sec:LindbladDerviation}. Throughout the paper we consider uniform Lindblad operators $L_i=S_i^z$ with a coupling strength $\gamma$. This type of coupling was first introduced in the study of dissipative two-state systems \cite{Leggett1987} and, as discussed below, is widely applicable for the description of environment-induced decoherence. For the simplest case of a single spin with $H=0$, the master equation \eqref{eq:master} predicts the typical exponential decay of off-diagonal density matrix elements, $\rho_{\uparrow,\downarrow}(t) \sim e^{-\gamma t/4}$, implying a rapid destruction of superpositions of states (quantum coherence). 

In this paper, we demonstrate using the example of spin-$1/2$ chains with internal interactions and the uniform couplings to the environment how the interplay between interaction and dissipation can fundamentally alter the decoherence behavior. In particular, (i) for the Heisenberg XXZ model in the thermodynamic limit, the coherence decay becomes algebraic instead of exponential, and (ii) for the transverse-field Ising model, the coherence decay remains exponential but the internal interactions can both facilitate and impede the decoherence in comparison to the non-interacting case. We provide quasi-exact numerical results using the time-dependent density matrix renormalization group method (tDMRG) \cite{Vidal2004,White2004,Daley2005,Zwolak2004-93} and explain both features on the basis of a perturbative treatment for the Liouville superoperator.

\section{Experimental realizations and applications}\label{sec:experiments}
Besides being of fundamental theoretical interest, the two dissipative models addressed in this paper are of broad experimental and technological relevance. We shortly mention a few examples.
A uniform coupling to the environment via Lindblad operators $L_i=S_i^z$ occurs for example naturally in quantum computer architectures \cite{DiVincenzo2000-48,Ladd2010-464} based on superconducting flux qubits (rf-SQUIDs) through fluctuations of the external magnetic flux \cite{Mooij1999-285,Yoshihara2006-97,Clarke2008-453}. Inductive coupling of flux qubits yields the Ising-type interaction $S^z_iS^z_j$ \cite{Majer2005-94,Johnson2011,Storcz2003-67}.
In ultracold atom systems \cite{Bloch2007} where both interaction and dissipation can be controlled, the corresponding Lindblad operators $L_i=n_i$ describe laser fluctuations and incoherent scattering of the laser light \cite{Anglin1997-79,Gerbier2010-82,Pichler2010}. 
With quantum dot spin-qubits \cite{Loss1998-57,Hanson2007-79}, one can implement both the transverse Ising model \cite{Shulman2012-336} and the Heisenberg model \cite{Loss1998-57}, where $L_i=S_i^z$ describes the effect of variations in the longitudinal nuclear magnetic field.

\section{Liouville spectrum}
Before addressing the two specific spin models, some general remarks are appropriate. First, as long as $L_i^\dag=L_i$ $\forall_i$, the maximally mixed state $\rho_0\propto \Id$ is always a steady state ($\partial_t{\rho}=0$) of Eq.~\eqref{eq:master}. For the models addressed in this paper, $\rho_0$ or restrictions of it to certain symmetry sectors are the unique steady states. Although all the initial states will eventually converge to such a steady state, the approach towards it is typically highly nontrivial and depends on the quantum many-body Hamiltonian. The dynamics is governed by the non-Hermitian superoperator $\hmc{L}$. Its eigenvalues $\lambda_\alpha$ have non-positive real-parts, $\Re \lambda_\alpha\leq 0$, and the steady state has the eigenvalue $\lambda_0=0$. We call
\begin{equation}\label{eq:gap}
	\Delta:=\min_{\alpha>0}\Re(- \lambda_\alpha)
\end{equation}
the spectral gap of the Liouville operator. If the gap is finite, the distance of the time-evolved state to the steady state will decrease exponentially with time, and $\Delta$ sets the corresponding relaxation rate. However, as we will see below, the many-body interactions in the system may qualitatively alter the dynamics by closing the gap of $\hmc{L}$ in the thermodynamic limit, which gives rise to a novel algebraic decoherence behavior.
\begin{figure}[t]
\includegraphics[width=0.89\linewidth]{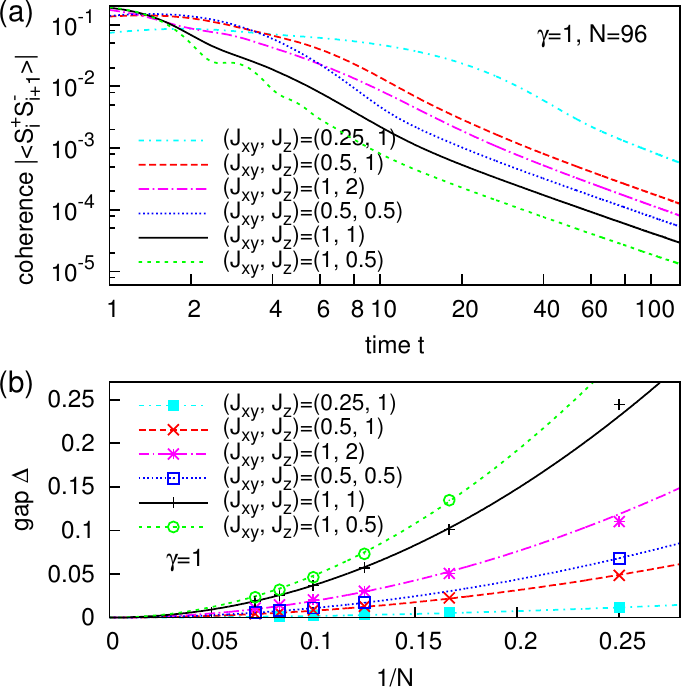}
\caption{\label{fig:XXZ}(a) Power-law decay of the off-diagonal density matrix element $C=\langle S_i^+S_{i+1}^-\rangle$ in the dissipative XXZ chain \eqref{eq:H_XXZ} of length $N=96$ for different $J_{xy}$, $J_z$, and fixed bath coupling $\gamma=1$, evaluated in the center of the chain.
(b) Finite-size scaling of the gap \eqref{eq:gap} of the Liouville superoperator $\hmc{L}$ for the open XXZ model, obtained by exact diagonalization (ED).}
\end{figure}

\section{Algebraic coherence decay in the open XXZ model}
First, let us consider the spin-$1/2$ XXZ chain
\begin{equation}\label{eq:H_XXZ}
	H_\text{XXZ}=\sum_i\left[ \frac{J_{xy}}2(S^+_iS^-_{i+1}+S^-_iS^+_{i+1})+J_z S_i^z S_{i+1}^z\right]
\end{equation}
uniformly coupled to the environment via the $z$-components of the spins, $L_i=S^z_i$. We study the time evolution of the system density matrix $\rho(t)$ based on the master equation \eqref{eq:master} with the initial state $\rho(0)=\ket{\Psi_0}\bra{\Psi_0}$ being the N\'{e}el state $\ket{\Psi_0}=\ket{\!\uparrow\downarrow\cdots\uparrow\downarrow}$.
In the absence of dissipation ($\gamma=0$), the time evolution for this setup has, for example, been studied in the context of quantum quenches \cite{Barmettler2009,Barmettler2010,Barthel2009-79}, where the long-time behavior decisively depends on $J_z/J_{xy}$. In this model, the total magnetization $\sum_i S^z_i$ is conserved. As a consequence, the off-diagonal element $\rho^i_{\uparrow,\downarrow}=\langle S_i^+\rangle$ of the single-site density matrices are strictly zero for all times and can not be used to monitor the decoherence. Instead, we can choose the off-diagonal term $C=\langle S_i^+S_{i+1}^-\rangle$ of the two-site density matrix to quantify the decoherence, where sites $i$ and $i+1$ are located in the center of the chain. For the simplest case of a two-site system ($N=2$), it is easy to show that, in the subspace of zero magnetization, the model can be mapped to the decoherence problem of a single spin subject to a transverse field and that the off-diagonal element $C$ decays exponentially (see appendix~\ref{sec:oneTwoSpins}).

In order to study the effects of the many-body correlations on the decoherence, we employ tDMRG \cite{Vidal2004,White2004,Daley2005,Zwolak2004-93}. As shown in Ref.\ \cite{Kliesch2011-107}, the propagator $\exp(\hmc{L}t)$ can be approximated by a circuit of two-site gates with an accuracy that is well-controlled in terms of the operationally relevant $(1\to 1)$-norm. Here, we specifically employ a fourth-order Trotter-Suzuki decomposition with a time step of size $\Delta t=0.125$. Starting from certain product states $\rho(0)$, the time-evolved states are obtained by applying the local Trotter gates and approximating $\rho(t)$ by matrix product states (MPS) \cite{Fannes1991,Rommer1997}. The essence of the DMRG procedure is to express $\rho(t)$ in every step of the simulation in a reduced Hilbert-Schmidt orthonormal operator basis $\{O^L_i\otimes O^R_i\}$ for a spatial splitting of the system into a left and a right part so that $\rho(t) = \sum_i \nu_i O^L_i\otimes O^R_i$.
Such a representation can always be obtained by singular value decomposition. The approximation consists in discarding all components $i$ with weights $\nu_i^2/\sum_j\nu_j^2$ below a certain threshold $\epsilon$ (between $10^{-10}$ and $10^{-12}$ in this work). One has to ensure convergence of the numerical results with respect to the truncation threshold $\epsilon$ and the system size $N$ in order to capture the physics of the thermodynamic limit. This is carried out in appendix~\ref{sec:cnvg}.

The evolution of the coherence $C(t)$ in open XXZ chains is shown in Fig.~\ref{fig:XXZ}. In the long-time limit, we find the coherence to decay algebraically, instead of exponentially, according to the power law
\begin{equation}
	C(t)\propto t^{-\eta} \quad\text{with}\quad \eta\approx 1.58.
\end{equation}
The exponent $\eta$ is in the studied parameter regime independent of the system parameters $J_{xy}$, $J_z$, and $\gamma$.
In general, an algebraic decay implies the absence of a characteristic time scale in the long-time dynamics. It results from the vanishing of the gap $\Delta$ of the Liouvillian $\hmc{L}$ in the thermodynamic limit. That this is indeed the case can be verified numerically by exact diagonalization as shown in Fig.~\ref{fig:XXZ}b.

To get a better understanding of this phenomenon, let us perform a second-order perturbative analysis to derive an effective Liouvillian $\hmc{L}_{\eff}$ for the limit $\gamma\gg |J_z|, |J_{xy}|$ of strong dissipation. The Liouvillian can be split into an unperturbed part $\hmc{L}_0\rho := -i[H_z,\rho] + \hmc{D}\rho$, where $H_{z}=J_{z}\sum_i S^z_iS^z_{i+1}$, and the perturbation $\hmc{L}_1\rho := -i[H_{xy},\rho]$ with $H_{xy}=\frac{J_{xy}}{2}\sum_i(S^+_iS^-_{i+1}+S^-_iS^+_{i+1})$.
The steady states of $\hmc{L}_0$ (eigenvalue $\lambda_0=0$) are $\rho^\vsigma_0=\ket{\vsigma}\bra{\vsigma}$, where $\ket{\vsigma}=\ket{\sigma_1\dots\sigma_N}$ are the $\{S^z_i\}$-eigenstates spanning the Hilbert space of the spin configurations with zero total magnetization. 
The effect of a small coupling $J_{xy}$ is to lift the degeneracy in the steady-state manifold through a superexchange process that leads us to an effective Liouvillian $\hmc{L}_{\eff}$, constrained to the subspace $\mathbb{H}$ spanned by the operators $\rho^\vsigma_0$. $\mathbb{H}$ is to be understood as a subspace of the vector space $\mathcal B(\mathscr{H})$ of linear operators on the Hilbert space $\mathscr{H}$, i.e., $\hmc{L}:\mathcal B(\mathscr{H})\to \mathcal B(\mathscr{H})$ and $\hmc{L}_{\eff}:\mathbb{H}\to\mathbb{H}$. One obtains
\begin{equation}\label{eq:Leff_gen}
	\hmc{L}_{\eff}=\hmc{P}\hmc{L}_1\frac{1}{\lambda_0-\hmc{L}_0}\hmc{L}_1\hmc{P}.
\end{equation}
$\hmc{P}$ is the projector onto the subspace $\mathbb{H}$. The intermediate states in the perturbation theory are of the form $\Lambda^{\vsigma\vsigma'}_1=\ket{\vsigma}\bra{\vsigma'}$, where $\ket{\vsigma'}=(S^+_iS^-_{i+1}+S^-_iS^+_{i+1})\ket{\vsigma}$ for some bond $(i,i+1)$. Their $\hmc{L}_0$-eigenvalues, needed to evaluate the denominator in Eq.~\eqref{eq:Leff_gen}, are $-\gamma$ or $-\gamma\pm iJ_z$, depending on $\vsigma$. However, the term $\pm iJ_z$ can be ignored as it represents an irrelevant contribution of order $1/\gamma^2$ to $\hmc{L}_{\eff}$.
The full calculation given in appendix~\ref{sec:perturbXXZ} shows that the matrix elements of the effective Liouvillian are identical with those of the ferromagnetic Heisenberg model
\begin{equation*}
	K=\frac{-J_{xy}^2}\gamma\sum_i \left[\frac{1}{2}(S^+_iS^-_{i+1}+S^-_iS^-_{i+1})+ S_i^z S_{i+1}^z-\frac{1}{4}\right]
\end{equation*}
in the sense that
\begin{equation}\label{eq:L_eff}
	\hmc{L}_{\eff}\ket{\vsigma}\bra{\vsigma}
	= -\sum_{\vsigma'} \bra{\vsigma'}K\ket{\vsigma}\cdot\ket{\vsigma'}\bra{\vsigma'}.
\end{equation}
As a consequence, at the level of the second-order perturbation theory, the gap \eqref{eq:gap} of the effective Liouvillian $\hmc{L}_{\eff}$ is that of the Heisenberg ferromagnet $K$. Its gap vanishes as $1/N^2$ due to the quadratic spin-wave dispersion around zero momentum and the $2\pi/N$ spacing of the quasi-momenta. This explains the quadratic behavior of the gaps $\Delta$ for the full model in Fig.~\ref{fig:XXZ}b.

\section{Decoherence in the open transverse Ising model}
\begin{figure}[tb]
\includegraphics[width=\linewidth]{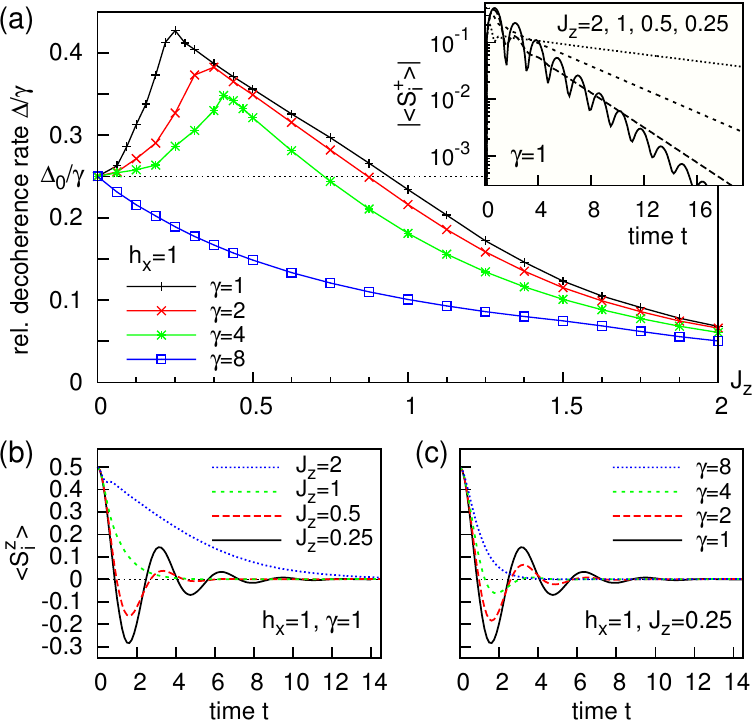}
\caption{\label{fig:delta}(a) The relative decoherence rate $\Delta/\gamma$ [Eq.~\eqref{eq:decohRate}] for the dissipative transverse Ising model \eqref{eq:H_tIsing} as a function of $J_z$ for different bath couplings $\gamma$. The rates were obtained by fitting the decay of the off-diagonal element $\rho_{\uparrow,\downarrow}^i=\langle S^+_i\rangle$ of the single-site density matrix, where site $i$ is in the middle of the chain. See the inset. (b) Time evolution of the magnetization $\langle S^z_i\rangle$ for fixed $\gamma=1$ and different $J_z$. (c) Dynamics of $\langle S^z_i\rangle$ for fixed $J_z=0.25$ and different $\gamma$. In all cases, the system contained $N=64$ sites.}
\end{figure}

A second paradigmatic example is the dissipative transverse-field Ising chain
\begin{eqnarray}\label{eq:H_tIsing}
   \nonumber H_\text{TI}&=&\sum_i( J_z \sigma_i^z \sigma_{i+1}^z-h_x
    \sigma_i^x)\\
    &=&\sum_i (4J_z S_i^z S_{i+1}^z-2h_x S_i^x)
\end{eqnarray}
with the interaction strength $J_z$, the transverse magnetic field $h_x$, and the Pauli matrices $\sigma^\alpha_i$. To study the interplay of interaction and dissipation, we set for simplicity $h_x=1$ and vary $J_z$ and the bath coupling $\gamma$. As Lindblad operators, we choose again $L_i=S^z_i$ and study the time evolution of the system density matrix based on Eq.~\eqref{eq:master} starting from a fully polarized state, i.e., $\rho(0)=\ket{\Psi'_0}\bra{\Psi'_0}$ with $\ket{\Psi'_0}=\ket{\!\!\uparrow\uparrow\cdots\uparrow\uparrow}$. Alternative initial states have also been checked. However, as we are foremost interested in the long-time behavior, the choice of the initial state is of minor importance. Let us first consider the noninteracting case with $J_z=0$ which reduces to the decoherence problem of a single spin subject to an external field. In this case, the off-diagonal element $\rho_{\uparrow,\downarrow}^i=\langle S^+_i\rangle$ of the single-site reduced density matrix decays exponentially as $\rho_{\uparrow,\downarrow}^i(t) \sim e^{-\Delta_0 t}$, where the decay rate is $\Delta_0=\gamma/4$ (as long as $\gamma\leq 8|h_x|$; see appendix~\ref{sec:oneTwoSpins}). For the interacting many-body system, one can use $|\rho_{\uparrow,\downarrow}^i(t)|$, with site $i$ in the middle of the chain, to monitor the coherence decay. In contrast to the situation for the open XXZ model, we find here that the coherence always decreases exponentially as shown in the inset of Fig.~\ref{fig:delta}a,
\begin{equation}\label{eq:decohRate}
    |\rho_{\uparrow,\downarrow}^i(t)|=|\langle S^+_i(t)\rangle|\sim e^{-\Delta t}.
\end{equation}
The decoherence rate (inverse relaxation time) $\Delta$ is determined by the interplay of the internal interaction and the dissipation.

For small $J_z$, the decoherence dynamics is well described by oscillations of exponentially decaying amplitude. For large $J_z$, $\rho_{\uparrow,\downarrow}^i(t)$ decays exponentially without oscillations. Similar behavior was found in the relaxation dynamics of the quenched XXZ chain without dissipation \cite{Barmettler2009,Barmettler2010,Barthel2009-79}. Now, let us turn to the question of whether the internal interaction facilitates or impedes the decoherence, i.e., whether the decoherence rate $\Delta$ is below or above that of the noninteracting case $J_z=0$ with $\Delta_0=\gamma/4$. As shown in Fig.~\ref{fig:delta}a, the answer to this question depends in an intricate manner on the values of $\gamma$ and $J_z$. In the presence of weak dissipation (small $\gamma$), the dependence of $\Delta$ on $J_z$ is non-monotonic. The interaction facilitates the environment-induced decoherence ($\Delta>\Delta_0$) for small $J_z$, whereas it impedes the decoherence ($\Delta<\Delta_0$) for large $J_z$. For sufficiently strong dissipation, the interaction always suppresses the decoherence.

A qualitative understanding of the fact that interaction and dissipation cooperate to enhance the decoherence in the case of small $J_z$ and $\gamma$ can be gained by analyzing the magnetization dynamics $\langle S^z_i(t)\rangle$ for different $J_z$ and $\gamma$, as shown in Figures~\ref{fig:delta}b and \ref{fig:delta}c. For small $J_z$ and $\gamma$, $\langle S^z_i\rangle$ is well described by an exponentially decaying oscillation. Generally, for open many-body systems, the long-range quantum correlations are usually destroyed during the long-time dissipative dynamics. As a consequence, the quantum entanglement between a single spin and the rest of the system is weak. This allows us to explain the above observations in a mean-field framework corresponding to the decoupling of the interaction term $S_i^z S_{i+1}^z\sim S^z_i\langle S^z\rangle$. On a qualitative level, the decoherence in the long-time limit can be understood as that of a single spin in a constant transverse field and a longitudinal field $2\langle S^z(t)\rangle$ due to its nearest neighbors. Figures \ref{fig:delta}b and \ref{fig:delta}c show that, for small $J_z$ and $\gamma$, the longitudinal field $\langle S^z(t)\rangle$ is quickly oscillating -- hence, playing a role similar to that of noise and thus accelerating the decoherence. Once the oscillations of $\langle S^z(t)\rangle$ vanish (large $J_z$ or $\gamma$), the decoherence is suppressed.

The second key observation, that strong interaction impedes the decoherence, can again be explained on the basis of a perturbative analysis, here in the limit of a weak magnetic field, $\gamma\gg |h_x|$. The field terms $\propto h_x$ of the Liouvillian are considered as a perturbation so that $\hmc{L}=\hmc{L}_0+\hmc{L}_1$ with
\begin{equation}
	\hmc{L}_0\rho=-i[H_{z},\rho]+\hmc{D}\rho\quad\text{and}\quad
	\hmc{L}_1\rho =-i[H_x,\rho],
\end{equation}
where $H_{z}=4J_z\sum_iS_i^z S_{i+1}^z$ and $H_x=-2h_x\sum_iS^x_i$. In the second-order perturbation theory, the eigenoperators of $\hmc{L}_0$ are similar to those in the treatment of the open XXZ model (now, the dynamics is not constrained to sectors of constant magnetization), but the intermediate states are different. Their $\hmc{L}_0$-eigenvalues are $-\gamma/2$ and $-\gamma/2\pm i4J_z$. The effective Liouville superoperator \eqref{eq:Leff_gen} is again of the form \eqref{eq:L_eff} and the effective Hamiltonian reads
\begin{equation}\label{eq:effTI}
K=\sum_i \left[\frac{\alpha+\alpha'}{4}-(\alpha-\alpha')S_{i-1}^zS_{i+1}^z\right]\left(\frac{1}{2}-S_i^x\right)
\end{equation}
in this case, where $\alpha=16h_x^2/\gamma$ and $\alpha'=4h_x^2\gamma/(\gamma^2/4+(4J_z)^2)$. On the basis of Eq.~\eqref{eq:effTI}, one can show that the gap \eqref{eq:gap} of the effective Liouvillian has for small $J_z$ a value $\lesssim (\alpha+\alpha')/4$. For sufficiently large $J_z$, the gap is given by $\alpha'$. The corresponding eigenstate of $K$ is the spin-wave-like state 
$\sum_j \ket{\!\!\uparrow^x\cdots\uparrow^x\downarrow^x_j\downarrow^x_{j+1}\uparrow^x\cdots\uparrow^x}$,
where $S^x_i\ket{\!\!\uparrow^x_i}=\frac{1}{2}\ket{\!\!\uparrow^x_i}$.
A detailed derivation is given in appendix~\ref{sec:perturbIsing}. So the gap decays as $1/J_z^2$, i.e., strong interaction impedes decoherence as we have found in the quasi-exact numerical analysis.

\section{Conclusion}
In summary, we have studied the long-time dynamics of open quantum spin systems, discovering that the quantum many-body effects can significantly change the nature of the environment-induced decoherence by either altering the exponential coherence decay to being algebraic or, alternatively, by increasing or decreasing the decay rate. Besides illustrative quasi-exact tDMRG simulations, we have explained those effects by a perturbative analysis.
The latter also indicates that these phenomena are certainly not limited to spin chains. Generically, algebraic coherence decay will occur for models where the eigenspace of the dissipative terms is highly degenerate, and this degeneracy is then broken through interactions within the system.
Another interesting direction for future investigations are driven-dissipative quantum many-body systems, where external forces drive the system far from equilibrium and the interplay between driving, dissipation, and internal interaction may give rise to further novel non-equilibrium phenomena.

\acknowledgments
We gratefully acknowledge discussions with U.\ Schollw\"ock and funding by the German Research Foundation through DFG FOR 801.

\appendix

\section{Derivations of the Markov-Lindblad master equation}\label{sec:LindbladDerviation}
The Lindblad master equation \eqref{eq:master}\newcommand{\eqMaster}{\eqref{eq:master}}
covers several different scenarios for (many-body) quantum systems coupled to some form of environment \cite{Weiss1999,Gardiner2004,Breuer2007,Alicki2007,Rivas2012}. Generically, all situations are covered, where the evolution of the system is due to a quantum dynamical semigroup or so-called infinitesimal divisible channels \cite{Lindblad1976-48,Wolf2008-279}. Let us shortly discuss some specific scenarios.

\subsection{Generalized Nakajima-Zwanzig equation}\label{sec:NZeq}
The Lindblad master equation can be obtained in the weak-coupling and singular-coupling regimes on the basis of the so-called generalized Nakajima-Zwanzig equation.
To this purpose, let us consider a system that is coupled to an environment as described by a Hamiltonian
\begin{equation*}
	H_\tot = H + H_B + V,
\end{equation*}
where $H$ is the system Hamiltonian, $H_B$ the environment Hamiltonian, and $V$ is the system-environment coupling. The time evolution of the total system is governed by the von Neumann equation $\partial_t \rho_\tot(t)=-i[H_\tot, \rho_\tot(t)]$. Following Ref.~\cite{Zwanzig1960-33} one can introduce the projector
\begin{equation}\label{eq:NZprojector}
	\hmc{P}\rho_\tot:=\Tr_B( \rho_\tot)\otimes \rho_B,
\end{equation}
go to the interaction picture with
\begin{align*}
	\tilde{\rho}_\tot(t)&:=e^{i(H+H_B)t}\rho_\tot(t)e^{-i(H+H_B)t}\quad\text{and}\\
	\tilde{V}(t)&:=e^{i(H+H_B)t}V e^{-i(H+H_B)t},
\end{align*}
and derive the following integral equation for the system density matrix
\begin{multline}\label{eq:NZ}
	\hmc{P}\tilde{\rho}_\tot(t)-\hmc{P}\tilde{\rho}_\tot(0)\\
	= \int_0^t\!\!\ud s\int_0^s\!\!\ud r \hmc{P}\hmc{V}(s)\hmc{\pi}(s,r)(1-\hmc{P})\hmc{V}(r)\hmc{P}\tilde{\rho}_\tot(r).
\end{multline}
In this expression, $\hmc{V}(t)\rho:=-i[\tilde{V}(t),\rho]$ and $\hmc{\pi}(t,r):=\mc{T}_+\exp\big(\int_r^t\ud s (1-\hmc{P})\hmc{V}(s)\big)$ with the time-ordering operator $\mc{T}_+$, and we have assumed that the initial state is a product state $\rho_\tot(0)=\rho(0)\otimes \rho_B$ with $[\rho_B,H_B]=0$ and that the coupling obeys $\Tr_B(V\rho_B)=0$ which can always be achieved by shifting terms between $V$ and $H_B$. See for example Refs.~\cite{Breuer2007,Rivas2012} for more details.

\subsection{Weak-coupling regime}\label{sec:weakCoupling}
The weak coupling regime can be analyzed by introducing a scalar coefficient $\alpha$ 
\begin{equation*}
	H_\tot \,\to\, H_\tot = H + H_B + \alpha V
\end{equation*}
and considering $\alpha\ll 1$. Expanding the right-hand side of the Nakajima-Zwanzig equation \eqref{eq:NZ} in $\alpha$ and truncating after the leading order (Born approximation), one obtains after the coordinate transformation $r\to s-r$
\begin{multline}\label{eq:BornSecular1}
	\hmc{P} \tilde{\rho}_\tot(t) - \hmc{P} \tilde{\rho}_\tot(0) \\
	\approx -\alpha^2\int_0^t\!\ud s\int_0^s\!\ud r \hmc{P}\hmc{V}(s)\hmc{V}(s-r)\hmc{P}\tilde{\rho}_\tot(s).
\end{multline}
One can now write the interaction in the form $V=\sum_n A_n\otimes B_n$ and express the operators $A_n$ using eigenstates $\ket{E}$ of the system Hamiltonian $H$ so that
\begin{gather}
	 A_n = \sum_\omega A_n(\omega) \quad\text{with} \label{eq:Aeigen}\\ \nonumber
	 A_n(\omega):=\mathop{\sum_{{E,E'}}}_{E-E'=\omega}\ket{E}\bra{E}A_n\ket{E'}\bra{E'}.
\end{gather}
Inserting this in Eq.~\eqref{eq:BornSecular1}, the integral over $r$ is traded for a sum over two energy differences $\omega,\omega'$ and the integrand acquires a phase factor $e^{i(\omega-\omega')s}$ as $\tilde{V}(t)=\sum_{\omega,n}e^{-i\omega t}A_n(\omega)\otimes \tilde{B}_n(t)=\sum_{\omega,n}e^{i\omega t}A^\dag_n(\omega)\otimes \tilde{B}^\dag_n(t)$. According to the Riemann-Lebesgue lemma, the quickly oscillating contributions with $\omega\neq \omega'$ vanish in the limit $\alpha\to 0$. Discarding all such terms, for a small but finite $\alpha$, corresponds to the so-called secular approximation which finally yields the Lindblad master equation \eqMaster{}.

\subsection{Singular-coupling regime}\label{sec:singularCoupling}
The opposing regime of singular coupling describes the situation where the system-environment coupling is much stronger than the system Hamiltonian but still much weaker than the environment Hamiltonian. It can be analyzed by introducing a scalar coefficient $\alpha$ 
\begin{equation*}
	H_\tot \,\to\, H_\tot = H + \frac{1}{\alpha^2}H_B + \frac{1}{\alpha} V
\end{equation*}
and considering $\alpha\ll 1$. Expanding Eq.~\eqref{eq:NZ} in $1/\alpha$, and using again a representation $V=\sum_n A_n\otimes B_n$ for the interaction, one obtains for the system density matrix in the interaction picture:
\begin{multline}\label{eq:Singular1}
	\tilde{\rho}(t) - \tilde{\rho}(0) = \int_0^t\!\ud s\int_0^s\!\ud r\\
	 \times\left(G_{mn}(s-r)\big[\tilde{A}_n(r)\tilde{\rho}(r),\tilde{A}_m(s)\big]+h.c.\right)+\mc{O}(\alpha^{-3})
\end{multline}
with the environment correlation function $G_{mn}(s-r)=\alpha^{-2}\Tr\big(\rho_B \tilde{B}_m \tilde{B}_n(r-s)\big)$. The higher-order terms in Eq.~\eqref{eq:Singular1} can be shown to vanish for $\alpha\to 0$ when $\rho_B$ in Eq.~\eqref{eq:NZprojector} is a Gaussian state \cite{Gorini1976-17b}. Expressing the operator $B_n$ using eigenstates of $H_B$ in analogy to Eq.~\eqref{eq:Aeigen}, $G_{mn}(s-r)$ attains the form of a Fourier integral and, along the lines of the Riemann-Lebesgue lemma, one can conclude that $G_{mn}(s-r)$ can be approximated by a delta function $\delta(s-r)$ for small $\alpha$, and one obtains the Lindblad master equation \eqMaster{}.

\subsection{Closed system with stochastic fields}\label{sec:stochastic}
One can realize Lindblad master equations with (closed) systems that comprise stochastic Hamiltonian terms, i.e., Hamiltonians of the form
\begin{equation}
	H_\tot=H+K(t)\quad\text{with}\quad
	K(t)=\sum_i\xi_i(t)F_i,
\end{equation}
where the $\xi_i(t)$ are scalar and $F_i^\dag=F_i$. The fields $\xi_i$ shall be independent and be generated by a Gaussian stochastic processes characterized by a continuous covariance function. In this case, all correlators of the stochastic fields are determined by the two-point correlators and we assume zero mean-values without loss of generality ($\overline{(\dots)}$ denotes the average over all random field configurations)
\begin{equation}\label{eq:rndField}
	\overline{\xi_i(t)}=0,\quad
	\overline{\xi_i(t)\xi_j(t')}=\gamma\delta_{ij}\delta_{\alpha}(t-t').
\end{equation}
The two-point correlations $\delta_\alpha(\Delta t)$ shall be peaked around $\Delta t=0$, have a width $\sim \alpha$, and obey $\int_0^\infty \ud s \delta_\alpha(s)=1/2$ such that $\lim_{\alpha\to 0}\delta_\alpha(\Delta t)=\delta(\Delta t)$. All higher-order correlators are fixed by Eq.~\eqref{eq:rndField} and can be evaluated using Wick's theorem
\begin{multline}\label{eq:Wick}
	 \overline{\xi_{i}(t)\xi_{i_1}(t_1)\dots \xi_{i_n}(t_n)}
	 =\sum_{k=1}^n \overline{\xi_{i}(t)\xi_{i_k}(t_k)}\\
	  \times \overline{\xi_{i_1}(t_1)\dots \xi_{i_{k-1}}(t_{k-1})\xi_{i_{k+1}}(t_{k+1})\dots \xi_{i_n}(t_n)}.
\end{multline}

With $\hmc{H}_\tot(t)\rho:=-i[H_\tot(t),\rho]$ and analogous definitions for $\hmc{H}$, $\hmc{K}(t)$, and $\hmc{F}_i$, the von Neumann equation $\partial_t\rho(t) = \hmc{H}_\tot(t) \rho(t)$ has the formal solution
\begin{equation*}
\rho(t)=\hat{\tau}(t,s)\rho(s)\quad\text{with}\quad
\hat{\tau}(t,r) = \mc{T}_+ e^{\int_r^t\ud s\hmc{H}_\tot(s)},
\end{equation*}
so that $\partial_t \hat{\tau}(t,r) = \hmc{H}_\tot(t)\hat{\tau}(t,r)$ and $\hat{\tau}(t,t)=1$. Let us prove in the following that the evolution of the averaged state $\overline{\rho(t)}=\overline{\hat{\tau}(t,s)}\rho(s)$ is governed by a Lindblad master equation. First, we introduce an interaction picture (denoted by tildes) with
\begin{equation}
	\tilde{\tau}(t,s) := e^{-\hmc{H}t}\hat{\tau}(t,s)
\end{equation}
so that
\begin{equation*}
 	\partial_t \tilde{\tau}(t,s) = \tmc{K}(t) \tilde{\tau}(t,s)\quad\text{with}\quad
 	\tmc{K}(t):=e^{-\hmc{H}t}\hmc{K}(t) e^{\hmc{H}t}.
\end{equation*}
With these definitions we have $\tilde{\tau}(t,r)=\mc{T}_+ e^{\int_r^t\ud s\tmc{K}(s)}$. Hence, the Dyson series for the propagator in the interaction picture is
\begin{equation}\label{eq:rndDyson1}
 	{\tilde{\tau}(t,s)} = \sum_{n=0}^\infty \int_s^t\ud t_1\dots \int_s^{t_{n-1}}\ud t_n
 	{\tmc{K}(t_1)\dots \tmc{K}(t_n)}.
\end{equation}
Taking the time derivative, employing the Wick theorem \eqref{eq:Wick}, plugging in Eq.~\eqref{eq:rndField}, and reordering integrals, the equation of motion for the averaged propagator is
\vspace{-1em}
\begin{widetext}
\begin{align*}
 	\partial_t\overline{\tilde{\tau}(t,0)} 
 	& = \sum_{n=0}^\infty \int_0^t\ud t_1\dots \int_0^{t_{n-2}}\!\!\ud t_{n-1}
 	\overline{\tmc{K}(t)\tmc{K}(t_1)\dots \tmc{K}(t_{n-1})}\\
 	& = \sum_{n=0}^\infty\sum_{k=1}^{n-1} \sum_{i,j}\int_0^t\ud t_1\dots \int_0^{t_{n-2}}\!\!\ud t_{n-1}
 	\overline{\xi_i(t)\xi_j(t_k)}\tmc{F}_i(t)\overline{\tmc{K}(t_1)\dots\tmc{K}(t_{k-1})\tmc{F}_j(t_k)\tmc{K}(t_{k+1})\dots \tmc{K}(t_{n-1})} \\
 	& = \sum_{m,n=0}^\infty \sum_{i,j} \int_0^t\ud s\int_s^t\ud t_1\dots \int_s^{t_{m-1}}\!\!\!\!\!\!\!\!\!\!\ud t_{m} \int_0^s\ud s_1\dots \int_0^{s_{n-1}}\!\!\!\!\!\!\!\!\!\!\ud s_{n} \gamma \delta_{ij}\delta_\alpha(t-s) \tmc{F}_i(t)\overline{\tmc{K}(t_1)\dots\tmc{K}(t_{m})\tmc{F}_j(s)\tmc{K}(s_1)\dots \tmc{K}(s_n)}.
\end{align*}
\end{widetext}
Reinserting Eq.~\eqref{eq:rndDyson1} and taking the limit $\alpha\to 0$, we finally obtain
\begin{align*}
 	\partial_t\overline{\tilde{\tau}(t,0)} 
 	& = \int_0^t\ud s \sum_i \gamma \delta_\alpha(t-s) \tmc{F}_i(t)\overline{\tilde{\tau}(t,s)\tmc{F}_i(s)\tilde{\tau}(s,0)}\\
 	& \stackrel{\alpha\to 0}{\longrightarrow}\underbrace{\frac{\gamma}{2}\sum_i\tmc{F}_i(t)\tmc{F}_i(t)}_{=:\tmc{D}(t)}\overline{\tilde{\tau}(t,0)}
\end{align*}
and, hence, in the Schr\"odinger picture
\begin{equation}\label{eq:rndLindbladMaster}
	\partial_t \overline{\hat{\tau}(t,0)} = (\hmc{H}+\hmc{D}) \overline{\hat{\tau}(t,0)}.
\end{equation}
The (dissipative) term has the form $\hmc{D}\rho = \frac{\gamma}{2}\sum_i(\hmc{F}_i)^2\rho = \gamma\sum_i\big(F_i\rho F_i-\frac{1}{2}\{F_i^2,\rho\}\big)$. So, Eq.~\eqref{eq:rndLindbladMaster} applied to $\rho(0)$ is just the Lindblad master equation \eqMaster{} for Hermitian Lindblad operators $L_i=F_i$.

\section{Perturbative treatment of the Liouvillians}\label{sec:perturb}
\subsection{Open spin-\texorpdfstring{$1/2$}{1/2} XXZ model with \texorpdfstring{$L_i=S^z_i$}{L=Sz}}\label{sec:perturbXXZ}
Let us again consider the spin-$1/2$ XXZ model \eqref{eq:H_XXZ} coupled to a Markovian environment according to the Lindblad master equation \eqMaster{} with uniform Lindblad operators $L_i=S^z_i$ and bath coupling strength $\gamma$. Based on a perturbative treatment, one can derive an effective Liouville superoperator $\hmc{L}_{\eff}$ for the limit of strong bath coupling $\gamma\gg |J_{xy}|,|J_z|$.
To this purpose, the Liouvillian is split into an unperturbed part $\hmc{L}_0$ and a perturbation $\hmc{L}_1$ such that
\begin{align}
	\hmc{L}&=\hmc{L}_0+\hmc{L}_1 \label{eq:XXZ_Lfull},\\
	\hmc{L}_0\rho &= -i[H_z,\rho] + \gamma\sum_i(S_i^z \rho S_i^z -\rho/4),\\
	\hmc{L}_1\rho &= -i[H_{xy},\rho],
\end{align}
where $H_{z}=J_{z}\sum_i S^z_iS^z_{i+1}$ and $H_{xy}=\frac{J_{xy}}{2}\sum_i(S^+_iS^-_{i+1}+S^-_iS^+_{i+1})$ as in Eq.~\eqref{eq:H_XXZ}. The Liouvillian conserves the total magnetization such that $\left[\sum_i S^z_i,\rho(t)\right]=0$ $\forall_t$. As in the main part of the paper, we choose the symmetry sector of zero magnetization for the following.

The steady states (eigenvalue $\lambda_0=0$) of the unperturbed part $\hmc{L}_0$ are of the form $\rho_0^\vsigma:=\ket{\vsigma}\bra{\vsigma}$ with $\{S^z_i\}$-eigenstates $\ket{\vsigma}=\ket{\sigma_1\dots\sigma_N}$ obeying  $S^z_i\ket{\vsigma}=\frac{\sigma_i}{2}\ket{\vsigma}$. Within the second-order perturbation theory, the resulting effective Liouvillian is given by Eq.~\eqref{eq:Leff_gen}
with $\hmc{P}$ projecting onto the space $\mathbb{H}$, spanned by the unperturbed steady states $\rho_0^\vsigma$. Starting from such a state, the intermediate states (after application of $\hmc{L}_1$) are of the form $\Lambda^{\vsigma\vsigma'}:=\ket{\vsigma}\bra{\vsigma'}$ such that $\ket{\vsigma'}=(S^+_iS^-_{i+1}+S^-_iS^+_{i+1})\ket{\vsigma}$ for some bond $(i,i+1)$. We need their $\hmc{L}_0$ eigenvalues to evaluate the denominator in Eq.~\eqref{eq:Leff_gen}. Computing $\hmc{L}_0 \Lambda^{\vsigma\vsigma'}$, one finds that the eigenvalues are $-\gamma$ or $-\gamma\pm iJ_z$ depending on the values of $\sigma_{i-1}$, $\sigma_{i}$ ($\sigma_{i+1}=-\sigma_{i}$), and $\sigma_{i+2}$. However, as $\gamma\gg |J_{xy}|,|J_z|$ and one only needs the effective Liouvillian up to first order in $1/\gamma$, the factor $1/(\lambda_0-\hmc{L}_0)$ in Eq.~\eqref{eq:Leff_gen} can for all cases be approximated by $1/\gamma$.
\begin{equation*}
	\frac{1}{\gamma\pm iJ_z}=\frac{1}{\gamma}+\mathcal{O}(\gamma^{-2})
\end{equation*}
With the terms $h_i:=\frac{J_{xy}}{2}(S^+_iS^-_{i+1}+S^-_iS^+_{i+1})$ in $H_{xy}$, the effective Liouvillian attains the form
\begin{align}
	\frac{1}{\gamma} \hmc{P}\hmc{L}_1\hmc{L}_1\rho_0^\vsigma 
	&= -\frac{1}{\gamma}\hmc{P} [H_{xy},[H_{xy},\rho_0^\vsigma ]]\nonumber\\
	&= -\frac{1}{\gamma}\sum_{i}\left( h_i^2 \rho_0^\vsigma + \rho_0^\vsigma h_i^2 -2 h_i \rho_0^\vsigma h_i \right).
\end{align}
Using that $h_i^2=\frac{J_{xy}^2}{2}\left(\frac{1}{4}-S^z_iS^z_{i+1}\right)$, one can read off the effective Hamiltonian $K$ that describes the action of the effective Liouvillian $\hmc{L}_{\eff}$ on the space $\mathbb{H}$ according to the definition \eqref{eq:L_eff}.
It turns out to be the isotropic Heisenberg ferromagnet with coupling constant $J_{xy}^2/\gamma$,
\begin{equation}\label{eq:XXZ_Heff}
	K = -\frac{J_{xy}^2}{\gamma}\sum_i\left[\frac{1}{2}(S^+_iS^-_{i+1}+S^-_iS^+_{i+1}) + S^z_iS^z_{i+1}-\frac{1}{4}\right].
\end{equation}

The spectra of $-\hmc{L}_{\eff}$ and $K$ are identical. The gap is decisive for the long-time dynamics and decoherence. To confirm and illustrate the perturbative analysis, we have done an exact diagonalization of the full model $\hmc{L}$ and the effective model $\hmc{L}_{\eff}$. As shown in Table~\ref{tab:XXZ_spectra_fullVsEff}, with increasing $\gamma$, the spectrum of $\hmc{L}_{\eff}$ converges indeed towards the corresponding eigenvalues of $\hmc{L}$. The exact gap of $\hmc{L}_{\eff}$ (finite only for finite system sizes $N$) is available, because spin waves with the well-known cosine dispersion relation are exact eigenstates of the isotropic Heisenberg ferromagnet \eqref{eq:XXZ_Heff}.
\begin{table}[htbp]
\begin{tabular}{|@{\quad}l@{\quad}||@{\quad}l@{\quad}||@{\quad}l@{\quad}|@{\quad}l@{\quad}|}
\hline
&&&\\[-1em]
$n$	& \quad$-\hmc{L}$& $\displaystyle\genfrac{}{}{-1pt}{}{-\hmc{L}_{\eff}}{\gamma=100}$	& $\displaystyle\genfrac{}{}{-1pt}{}{-\hmc{L}_{\eff}}{\gamma=10}$\\[0.7em]
\hline
0	& 0		& 0		& 0\\
1	& 0.292893	& 0.292877	& 0.291271\\
2	& 0.633975	& 0.633925	& 0.629092\\
3	& 1.000000	& 0.999994	& 0.999369\\
4	& 1.707107	& 1.707042	& 1.700617\\
5	& 2.366025	& 2.365987	& 2.362134\\
\hline
\end{tabular}
\caption{\label{tab:XXZ_spectra_fullVsEff}
Spectrum of the effective model $\hmc{L}_{\eff}$ [Eq.~\eqref{eq:XXZ_Heff}] compared to the corresponding eigenvalues of the full Liouvillian [Eq.~\eqref{eq:XXZ_Lfull}] for a XXZ chain of $N=4$ sites with open boundary conditions, $J_{xy}=0.5$, and $J_z=1$. The eigenvalues are given in units of the effective coupling ${J_{xy}^2}/{\gamma}$.}
\end{table}

\subsection{Open transverse Ising model with \texorpdfstring{$L_i=S^z_i$}{L=Sz}}\label{sec:perturbIsing}
This appendix presents, in full detail, the perturbation theory for the spin-$1/2$ Ising model in a transverse field \eqref{eq:H_tIsing} coupled to a Markovian environment according to the Lindblad master equation \eqMaster{} with $L_i=S^z_i$ and bath coupling strength $\gamma$. Using a second-order perturbation theory, one can derive an effective Liouville superoperator $\hmc{L}_{\eff}$ for the limit of a weak magnetic field $\gamma\gg |h_x|$.
To this purpose, the Liouvillian is split into an unperturbed part $\hmc{L}_0$ and a perturbation $\hmc{L}_1$ such that
\begin{align}
	\hmc{L}&=\hmc{L}_0+\hmc{L}_1 \label{eq:tIsing_Lfull},\\
	\hmc{L}_0\rho &= -i[H_z,\rho] + \gamma\sum_i(S_i^z \rho S_i^z -\rho/4),\\
	\hmc{L}_1\rho &= -i[H_{x},\rho],
\end{align}
where $H_{z}=4 J_{z}\sum_i S^z_iS^z_{i+1}$ and $H_{x}=-2h_x\sum_i S^x_i$ as in Eq.~\eqref{eq:H_tIsing}.

As in the perturbative analysis of the XXZ model, the steady states (eigenvalue $\lambda_0=0$) of the unperturbed part $\hmc{L}_0$ are of the form $\rho_0^\vsigma:=\ket{\vsigma}\bra{\vsigma}$ with $\{S^z_i\}$-eigenstates $\ket{\vsigma}$ obeying  $S^z_i\ket{\vsigma}=\frac{\sigma_i}{2}\ket{\vsigma}$. The effective Liouvillian is given by Eq.~\eqref{eq:Leff_gen} with $\hmc{P}$ projecting onto the space $\mathbb{H}$ that is spanned by the unperturbed steady states $\rho_0^\vsigma$. The differences to the calculation for the XXZ model in appendix~\ref{sec:perturbXXZ} are that, here, the total magnetization is not conserved and that the factor $1/(\lambda_0-\hmc{L}_0)$ in Eq.~\eqref{eq:Leff_gen} is now nontrivial. Starting from one of the unperturbed steady states, the intermediate states (after application of $\hmc{L}_1$) are of the form $\Lambda^{\vsigma\vsigma'}:=\ket{\vsigma}\bra{\vsigma'}$ such that $\ket{\vsigma'}=2S^x_i\ket{\vsigma}$ for some site $i$. Computing $\hmc{L}_0 \Lambda^{\vsigma\vsigma'}$, one finds that the $\hmc{L}_0$ eigenvalues of the intermediate states are $-\gamma/2-i(E_{\vsigma}-E_{\vsigma'})$. Here, $E_{\vsigma}:=\bra{\vsigma}H_z\ket{\vsigma}$ and $E_{\vsigma}-E_{\vsigma'}=0$ or $\pm4 J_z$ depending on the values of $\sigma_{i-1}$, $\sigma_{i}$, and $\sigma_{i+1}$. With the definition of the state $\ket{\vsigma,i}:=2S_i^x\ket{\vsigma}$ and the corresponding eigenvalue of $H_z$ labeled by $E^i_{\vsigma}$, the effective Liouvillian \eqref{eq:Leff_gen} takes the form
\begin{align}
	\hmc{L}_{\eff}\ket{\vsigma}\bra{\vsigma} &= \hmc{P}\hmc{L}_1\frac{1}{\lambda_0-\hmc{L}_0}\hmc{L}_1\ket{\vsigma}\bra{\vsigma}\nonumber\\
	&= i\hmc{P}\hmc{L}_1\sum_i \left(\frac{h_x\ket{\vsigma,i}\bra{\vsigma}}{\gamma/2+i(E^i_{\vsigma}-E_{\vsigma})} - h.c. \right)\nonumber\\
	&= -\gamma h_x^2\sum_i\frac{\ket{\vsigma}\bra{\vsigma}-\ket{\vsigma,i}\bra{\vsigma,i}}{(\gamma/2)^2 + (E^i_{\vsigma}-E_{\vsigma})^2}.
\end{align}
Employing $|E^i_{\vsigma}-E_{\vsigma}|=2|J_z(\sigma_{i-1}+\sigma_{i+1})|$, one can read off the effective Hamiltonian $K$ that describes the action of the effective Liouvillian $\hmc{L}_{\eff}$ on the space $\mathbb{H}$ according to the definition \eqref{eq:L_eff} and finds
\begin{align}
	K =&\quad\,\,\alpha \sum_i\left(\frac{1}{4}-S^z_{i-1}S^z_{i+1}\right)\left(\frac{1}{2}-S_i^x\right)\nonumber\\
	&+\alpha' \sum_i\left(\frac{1}{4}+S^z_{i-1}S^z_{i+1}\right)\left(\frac{1}{2}-S_i^x\right)\nonumber\\
	=& \sum_i\left(\frac{\alpha+\alpha'}{4}-(\alpha-\alpha')S^z_{i-1}S^z_{i+1}\right)\left(\frac{1}{2}-S_i^x\right)
	\label{eq:tIsing_Heff}
\end{align}
with
\begin{equation}
	\alpha=\frac{\gamma(2h_x)^2}{(\gamma/2)^2}
	\quad\text{and}\quad
	\alpha'=\frac{\gamma(2h_x)^2}{(\gamma/2)^2+(4J_z)^2}.
\end{equation}

We are particularly interested in the gap \eqref{eq:gap} of $\hmc{L}_{\eff}$ which is given by that of $K$ and determines the decoherence. Let $\ket{\!\!\uparrow^x_i}$ and $\ket{\!\!\downarrow^x_i}$ denote the normalized eigenstates of $S^x_i$ with eigenvalues $\pm 1/2$. In this notation, the fully polarized state 
\begin{equation}
	\ket{\psi_0}=\ket{\!\uparrow^x\uparrow^x\cdots\uparrow^x}
\end{equation}
is the eigenstate of $K$ with eigenvalue $\lambda_0=0$. As $K$ is positive-semidefinite ($K\succeq 0$), $\ket{\psi_0}$ is its ground state. Note that the counterpart of $\ket{\psi_0}$ in $\mathbb{H}$ is the maximally mixed state $\Id=\sum_{\vsigma}\ket{\vsigma}\bra{\vsigma}$ which is also the exact steady state of $\hmc{L}$.

In order to determine or bound the gap, the first intuition may be to flip a single spin, i.e., to consider states
\begin{equation} \label{eq:tIsing_flipState}
	\ket{\phi_1^i}:=\ket{\!\uparrow^x\uparrow^x\cdots\uparrow^x\downarrow^x_i\uparrow^x\cdots\uparrow^x}.
\end{equation}
They provide the upper bound
\begin{equation} \label{eq:tIsing_flipGap}
	\bra{\phi_1^i} K \ket{\phi_1^i}= \frac{\alpha+\alpha'}{4}
\end{equation}
to the gap. The same bound pertains for any superposition of the states $\ket{\phi_1^i}$.
One can also construct exact spin-wave-type eigenstates $\ket{\tilde\psi_1^k}$ where a pair of spin defects is bound and travels through the lattice.
\begin{align}
	\ket{\psi_1^j}&:=\ket{\!\uparrow^x\uparrow^x\cdots\uparrow^x\downarrow^x_j\downarrow^x_{j+1}\uparrow^x\cdots\uparrow^x}\\
	\ket{\tilde\psi_1^k}&:=N^{-1/2}\sum_j e^{-ikj} \ket{\psi_1^j} \label{eq:tIsing_waveState}
\end{align}
From $K\ket{\psi_1^j}=\frac{\alpha+\alpha'}{2}\ket{\psi_1^j}-\frac{\alpha-\alpha'}{4}(\ket{\psi_1^{j-1}}+\ket{\psi_1^{j+1}})$ follows the eigenequation
\begin{equation}
	 K\ket{\tilde\psi_1^k}=\left(\frac{\alpha+\alpha'}{2}-\frac{\alpha-\alpha'}{2}\cos(k)\right)\ket{\tilde\psi_1^k}.
\end{equation}
The minimum eigenvalue $\alpha'$ is obtained for $k=0$.
\begin{figure*}[tb]
\centering
\includegraphics[width=0.47\linewidth]{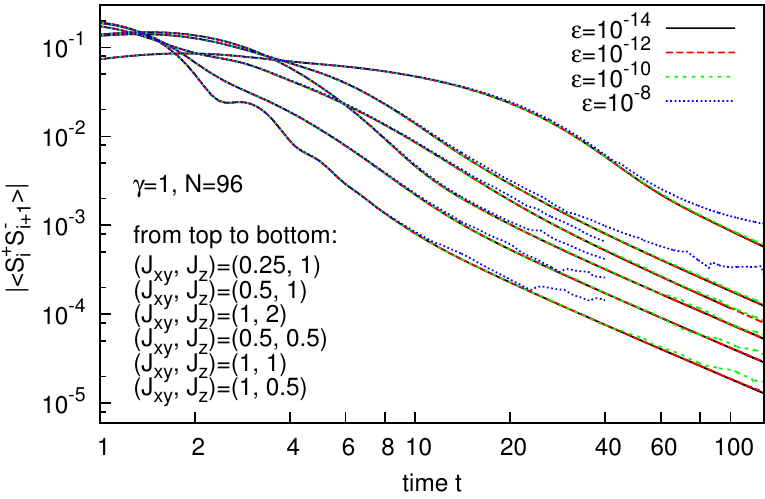}\quad\quad
\includegraphics[width=0.47\linewidth]{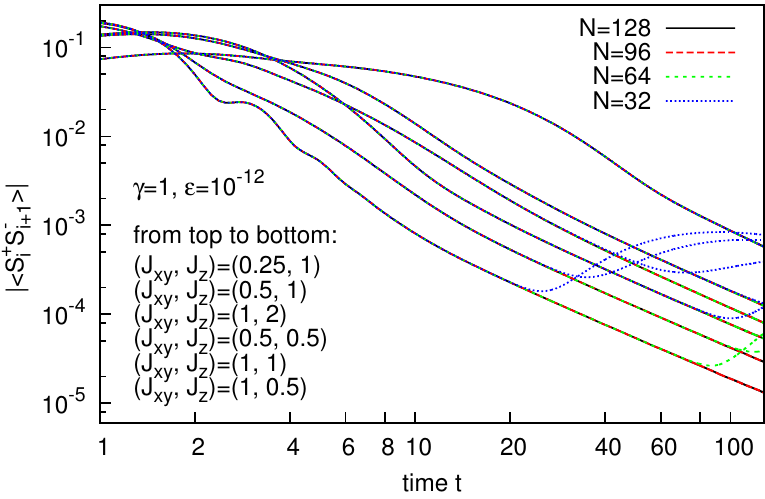}\\[1em]
\caption{\label{fig:XXZ_cnvg}Convergence analysis for the open XXZ model \eqref{eq:XXZ_Lfull} at the example of the expectation value $\langle S^+_iS^-_{i+1}\rangle$ with different $J_{xy}$ and $J_z$, and fixed bath coupling $\gamma=1$, where site $i$ is in the middle of the chain. Left: variation of the DMRG truncation threshold $\epsilon$ for the system size $N=96$. Right: variation of the system size for the truncation threshold $\epsilon=10^{-12}$.}
\end{figure*}
\begin{figure*}[tb]
\centering
\includegraphics[width=0.47\linewidth]{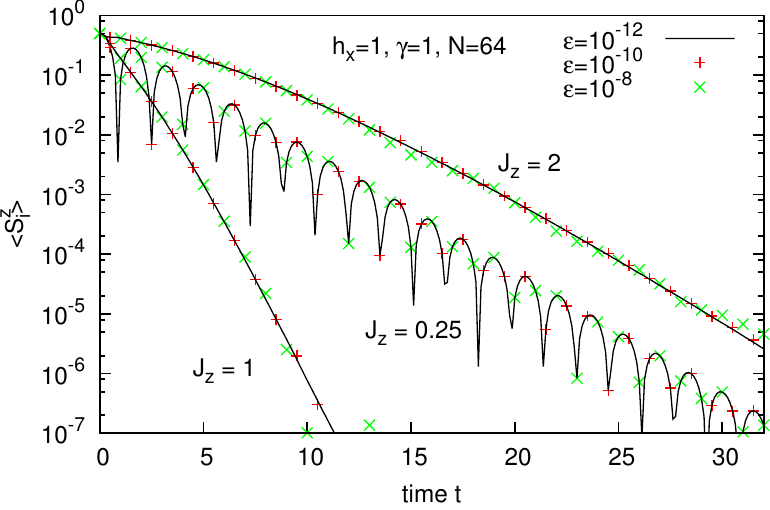}\quad\quad
\includegraphics[width=0.47\linewidth]{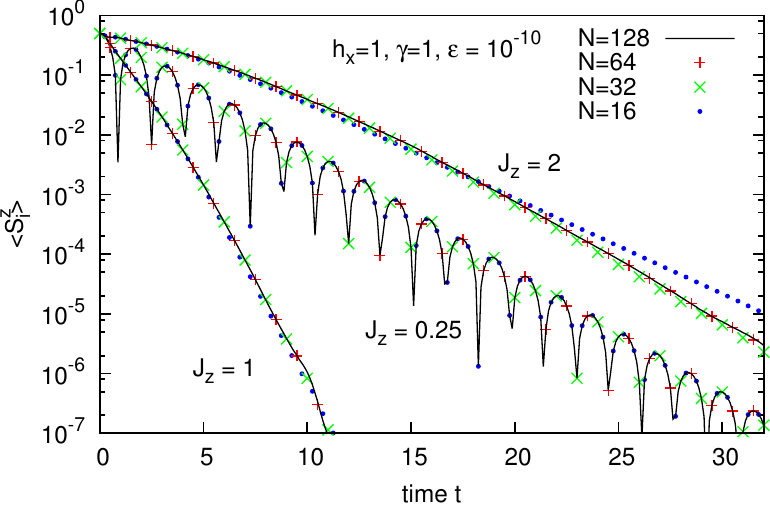}\\[1em]
\caption{\label{fig:tIsing_cnvg}Convergence analysis for the open transverse-field Ising model \eqref{eq:tIsing_Lfull} at the example of the magnetization $\langle S^z_i\rangle$
with different $J_z$, fixed bath coupling $\gamma=1$, and field $h_x=1$, where site $i$ is in the middle of the chain. Left: variation of the DMRG truncation threshold $\epsilon$ for the system size $N=64$. Right: variation of the system size for the truncation threshold $\epsilon=10^{-10}$.}
\end{figure*}

Using exact diagonalization (ED), we have asserted that the first excited state is similar to $\ket{\phi_1^i}$ [Eq.~\eqref{eq:tIsing_flipState}] for $\alpha'/\alpha\simeq 1$ (small $J_z$) with a gap $\lambda_1\sim\frac{\alpha+\alpha'}{4}$. For sufficiently small $\alpha'/\alpha$ (sufficiently large $J_z$), $\ket{\tilde\psi_1^k}$ [Eq.~\eqref{eq:tIsing_waveState}] becomes the first excited state with the gap $\lambda_1=\alpha'$. In contrast to the open XXZ model discussed in appendix~\ref{sec:perturbXXZ}, the gap for the thermodynamic limit is hence finite. Consequently, the coherence decay is always exponential in this model as described in the main text.

\section{One and two spins with \texorpdfstring{$L_i=S^z_i$}{L=Sz}}\label{sec:oneTwoSpins}
For completeness and in order to have a consistency check for the the perturbative analysis presented above in appendix~\ref{sec:perturbIsing}, let us shortly discuss the decoherence for a single spin-$1/2$ in a transverse magnetic field. The Liouville master equation for the Hamiltonian $H=-2h_x S^x$ and the Lindblad operator $L=S^z$ reads
\begin{equation}\label{eq:singleSpin_L}
	\partial_t\rho = i[2h_x S^x,\rho]+\gamma(S^z\rho S^z-\rho/4).
\end{equation}
Parameterizing the density, matrix as $\rho=\tsmat{a&b+ic\\b-ic&1-a}$, we have
\begin{equation*}
	\partial_t\rho
	= ih_x\tmat{-2ic&1-2a\\2a-1&2ic}
	  -\frac{\gamma}{2}\tmat{0&b+ic\\b-ic&0},
\end{equation*}
giving the three equations
\begin{equation*}
	\partial_t a = 2h_x c,\quad
	\partial_t b = -\frac{\gamma}{2} b,\quad
	\partial_t c = h_x(1-2a)-\frac{\gamma}{2}c.
\end{equation*}
So $b$ decays as $b(t)=b(0)e^{-\gamma t/2}$. The first and third equations imply $\partial_t^2c =-4 h_x^2c-\frac{\gamma}{2}\partial_t c$ with the solution $c(t)=c_+e^{i\omega_+ t}+c_-e^{i\omega_- t}+c.c.$ and
\begin{equation}\label{eq:singleSpin_omega}
	\omega_\pm = \frac{i\gamma}{4}\pm\sqrt{4h_x^2-\left(\frac{\gamma}{4}\right)^2}.
\end{equation}
For $\gamma\leq 8 |h_x|$, the coherence hence decays in the long-time limit as $\sim e^{-\gamma t/4}$. For large $\gamma$ however ($\gamma>8 |h_x|$), the square root in Eq.~\eqref{eq:singleSpin_omega} becomes imaginary and, thus, alters the decoherence rate from $\gamma/4$ to $\gamma/4-\sqrt{(\gamma/4)^2-4h_x^2}=8 h_x^2/\gamma+\mathcal{O}(h_x^4/\gamma^3)$. This can be compared with the perturbative upper bound $(\alpha+\alpha')/4$ [Eq.~\eqref{eq:tIsing_flipGap}] for the gap of the transverse-field Ising model as derived in appendix~\ref{sec:perturbIsing}. For $J_z=0$, this bound has indeed the same value found here, $(\alpha+\alpha')/4=8 h_x^2/\gamma$. So the two computations are consistent.

In the main text, we also mention the decoherence for two spins with zero total magnetization ($S^z_1+S^z_2=0$) evolving according to the Lindblad master equation with the XXZ interaction $H'=\frac{J_{xy}}{2}(S_1^+S_2^-+S_1^-S_2^+) +J_z S_1^zS_2^z$ and the dissipative term $\hmc{D}'\rho' = \gamma'(S^z_1\rho' S_1^z+S^z_2\rho' S_2^z-\rho'/2)$. This model maps to the single spin in a transverse field \eqref{eq:singleSpin_L} if we associate the state $\ket{\!\!\uparrow\downarrow}$ with the spin-up state, and $\ket{\!\!\downarrow\uparrow}$ with the spin-down state. The resulting single-spin dynamics is governed by the Hamiltonian $H=J_{xy}S^x-J_z/4$ and the dissipative term $\hmc{D}\rho = 2\gamma'(S^z\rho S^z-\rho/4)$. So the equation of motion is just Eq.~\eqref{eq:singleSpin_L} with transverse field $h_x=-J_{xy}/2$ and bath coupling $\gamma=2\gamma'$. Hence, for $\gamma'\leq 2 |J_{xy}|$, the decoherence rate is $\gamma'/2$. As is shown in the main part of the paper, the decoherence changes from this exponential behavior to a power-law decay in the thermodynamic limit of the corresponding many-particle system.

\section{Convergence of the tDMRG results}\label{sec:cnvg}
The numerical investigations are based on the time-dependent density matrix renormalization group (tDMRG) \cite{Vidal2004,White2004,Daley2005}. We use a Trotter-Suzuki decomposition of the propagator $\exp(\hmc{L}t)$ \cite{Kliesch2011-107} for which the error is of fifth order in the time step $\Delta t=0.125$. As described in the main text, an essential part of the algorithm is to truncate small-weight components in the singular value decomposition of the density matrix $\rho(t)$. This is necessary in order to bound the computation cost. The corresponding truncation threshold $\epsilon$ determines the accuracy and the computation cost of the simulation. One has to ensure convergence of the numerical results with respect to $\epsilon$ and with respect to the system size $N$, as we are interested in the physics of the thermodynamic limit $N\to\infty$.

Figure~\ref{fig:XXZ_cnvg} shows the analysis for the dissipative XXZ model \eqref{eq:XXZ_Lfull}. The results are sufficiently converged for a truncation threshold of $\epsilon=10^{-12}$ and the system size $N=96$. The simulations presented in the main text were done with these parameters.
Figure~\ref{fig:tIsing_cnvg} shows the analysis for the dissipative transverse-field Ising model \eqref{eq:tIsing_Lfull}. A truncation threshold of $\epsilon=10^{-10}$ and the system size $N=64$ are sufficient and were used for the simulations presented in the main text.

\bibliographystyle{prsty}

\end{document}